\documentclass{appolb}
\usepackage{epsfig}

\def\lsim{\mathrel{\rlap{\lower4pt\hbox{\hskip1pt$\sim$}}
    \raise1pt\hbox{$<$}}}                % less than or approx. symbol
\def\gsim{\mathrel{\rlap{\lower4pt\hbox{\hskip1pt$\sim$}}
    \raise1pt\hbox{$>$}}}                % greater than or approx. symbol
    
\begin{document}
\eqsec  % uncomment this line to get equations numbered by (sec.num)
\title{QED Thermodynamics at Intermediate Coupling%
}
\author{Michael Strickland
\address{Department of Physics, Gettysburg College, Gettysburg, PA 17325, USA}
\\$\;$\\
Nan Su
\address{Frankfurt Institute for Advanced Studies,  D-60438 Frankfurt 
am Main, Germany}
\\$\;$\\
Jens O. Andersen
\address{Department of Physics, Norwegian University of Science
and Technology, H{\o}gskoleringen 5, N-7491 Trondheim, Norway}
}
\maketitle
\begin{abstract}
We discuss reorganizing finite temperature perturbation theory using hard-thermal-loop (HTL) perturbation
theory in order to improve the convergence of successive perturbative approximations to the free energy of
a gauge theory.  We briefly review the history of the technique and present new results for the three-loop
HTL-improved approximation for the free energy of QED.  We show that the hard-thermal-loop perturbation
reorganization improves the convergence of the successive approximations to the QED free energy at intermediate
coupling, $e \sim 2$.  The reorganization is gauge invariant by construction, and  due to cancellation among various contributions, one can obtain a completely analytic result for the resummed thermodynamic potential at three loops.
\end{abstract}
  
\section{Introduction}

In the early 1990s the free energy of a massless scalar field theory was calculated
to order $g^4$ in Refs.~\cite{wrong,AZ-95}.  This was quickly followed by similar 
calculations in QED \cite{qed4} and QCD \cite{AZ-95}.
The scalar, QED, and QCD free energies to order $g^5$ were then obtained 
in Refs.~\cite{singh,ea1}, Refs.~\cite{parwani,Andersen} and
Refs.~\cite{KZ-96,BN-96}, respectively.
Recent results have extended the calculation of the QCD free energy
by determining the coefficient of the $g^6 \log(g)$ contribution
\cite{Kajantie:2002wa}.  For massless scalar $\phi^4$ the
perturbative free energy is now known to order $g^6$ \cite{Gynther:2007bw}
and $g^8\log(g)$ \cite{Andersen:2009ct}.

However, the resulting weak-coupling
approximations, truncated order-by-order in the 
coupling constant, are poorly 
convergent
unless the coupling constant is extremely small. For example, simply comparing
the magnitude of low-order contributions to the $N_f=3$ QCD
free energy one finds that the $g_s^3$ contribution is smaller than the $g_s^2$
contribution only for $g_s \lsim 0.9$ ($\alpha_s \lsim 0.07$).  This is a
troubling situation
since at phenomenologically accessible temperatures near the critical 
temperature 
for the QCD deconfinement phase transition, the strong coupling constant
is on the order of $g_s \sim 2$.

%%%%%%%%%%%%%%%%%%%%%%%%%%%%%%%%%%%%%%%%%%%%%%%%%%%%%%%%%%%%%%%%
\begin{figure}[t]
\begin{center}
\vspace{1cm}
\includegraphics[width=9.5cm]{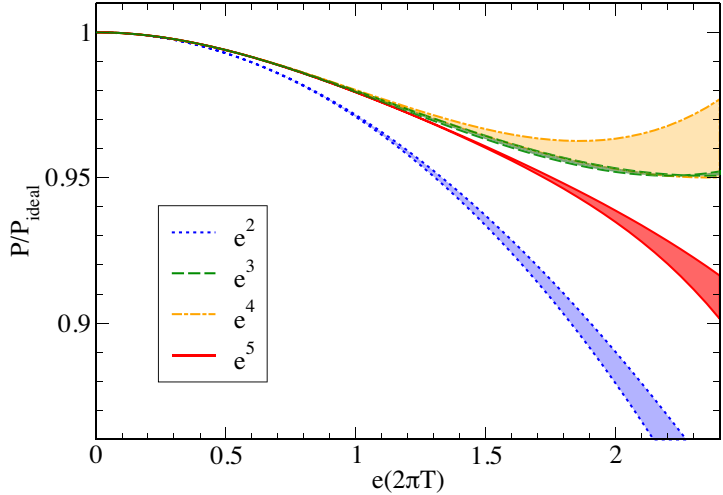}
\end{center}
\vspace{-5mm}
\caption{Successive perturbative approximations to the QED pressure 
(negative of the free energy).  
Each band corresponds to a
truncated weak-coupling expansion accurate to order $e^2$, $e^3$, $e^4$, and 
$e^5$, respectively.  
Shaded bands correspond to variation of the renormalization scale $\mu$ 
between $\pi T$ and $4 \pi T$.}
\label{fig:pertpressure}
\end{figure}
%%%%%%%%%%%%%%%%%%%%%%%%%%%%%%%%%%%%%%%%%%%%%%%%%%%%%%%%%%%%%%%%

The poor convergence of finite-temperature perturbative expansions of 
the free energy
is not limited to QCD.  The same behavior can be seen in weak-coupling
expansions
in scalar field theory \cite{spt,Andersen:2008bz} and QED \cite{parwani}.  
In Fig.~\ref{fig:pertpressure} we show the successive perturbative 
approximations
to the QED free energy.  As can be seen from this figure, at couplings larger 
than
$e \sim 1$ the QED weak-coupling approximations also exhibit poor 
convergence.  
For this reason a concerted effort
has been put forth to find a reorganization of finite-temperature 
perturbation theory which converges at phenomenologically relevant
couplings.

There are several ways of systematically 
reorganizing the perturbative expansion to improve its convergence 
and the various approaches have been reviewed 
in Refs.~\cite{birdie,kraemmer, review}. Here we will 
describe recent advances in the application of 
hard-thermal-loop perturbation theory
(HTLpt) \cite{htl1,AS-01,fermions,htl2,aps1}.  The HTLpt 
method is inspired by variational perturbation 
theory~\cite{yuk,steve,kleinert,deltaexp} and is a 
a gauge-invariant
extension of scalar screened perturbation theory 
(SPT)~\cite{K-P-P-97,CK-98,spt,Andersen:2008bz}.
The basic idea of the technique is to add and subtract an effective mass term from 
the bare Lagrangian and to associate the added piece with the 
free Lagrangian and the subtracted piece with the interactions.
However, in gauge theories, one cannot simply add and subtract a local mass
term since this would violate gauge invariance.
Instead one adds and subtracts a HTL 
improvement term which modifies the propagators and
vertices in such a way that the framework is manifestly gauge-invariant.
The free part of the Lagrangian then includes the HTL self-energies and 
the remaining terms are treated as perturbations.

In this brief proceedings review we present results of a calculations of
the QED free energy (pressure) to three-loop order in HTLpt based on
the work detailed in Ref.~\cite{Andersen:2009tw}.  
As we will show,
the next-to-leading order (NLO) and next-to-next-to-leading order (NNLO)
HTLpt resummed QED free energy give
approximations which show improved convergence for couplings as large
as $e \sim 2.5$ (see Fig.~\ref{fig:NLONNLO}).  In addition, we compare 
our results to those obtained using the 2PI $\Phi$-derivable approach 
\cite{phijm}
and show that at three loops the agreement between the HTLpt and 
$\Phi$-derivable approaches is quite good.

\section{Formalism}

\label{HTLpt}

The Lagrangian density for massless QED in Minkowski space is
\begin{eqnarray}\nonumber
{\cal L}_{\rm QED}&=&
-{1\over4}F_{\mu\nu}F^{\mu\nu}
+i \bar\psi \gamma^\mu D_\mu \psi 
\\&&\hspace{9mm} 
+{\cal L}_{\rm gf}
+{\cal L}_{\rm gh}
+\Delta{\cal L}_{\rm QED}\;.
\label{L-QED}
\end{eqnarray}
%
%\checked{mnj}
Here the field strength is 
$F^{\mu\nu}=\partial^{\mu}A^{\nu}-\partial^{\nu}A^{\mu}$
and the covariant derivative is $D^{\mu}=\partial^{\mu}+ieA^{\mu}$.
The ghost term ${\cal L}_{\rm gh}$ depends on the gauge-fixing term
${\cal L}_{\rm gf}$.  We will use dimensional regularization with a
renormalization scale $\mu$ and covariant gauge fixing such that
the ghost terms decouple.

Hard-thermal-loop perturbation theory is a reorganization
of the perturbation
series for thermal gauge theories. In the case of QED, 
the Lagrangian density is written as
\begin{eqnarray}
{\cal L}= \left({\cal L}_{\rm QED}
+ {\cal L}_{\rm HTL} \right) \Big|_{e \to \sqrt{\delta} e}
+ \Delta{\cal L}_{\rm HTL}\;.
\label{L-HTLQCD}
\end{eqnarray}
%\checked{mnj}
The HTL improvement term is
\begin{eqnarray}
{\cal L}_{\rm HTL}=-{1\over2}(1-\delta)m_D^2 %{\rm Tr}\left(
F_{\mu\alpha}\left\langle {y^{\alpha}y^{\beta}\over(y\cdot\partial)^2}
	\right\rangle_{\!\!y}F^{\mu}_{\;\;\beta}
%\right)
	\nonumber \\
         +(1-\delta)\,i m_f^2 \bar{\psi}\gamma^\mu 
\left\langle {y^{\mu}\over y\cdot D}
	\right\rangle_{\!\!y}\psi
	\, ,
\label{L-HTL}
\end{eqnarray}
%\checked{mnj}
where $y^{\mu}=(1,\hat{{\bf y}})$ is a light-like four-vector,
and $\langle\ldots\rangle_{ y}$
represents an average over the directions
of $\hat{{\bf y}}$.
The term~(\ref{L-HTL}) has the form of the effective Lagrangian
that would be induced by
a rotationally-invariant ensemble of charged sources with infinitely high
momentum. The parameter $m_D$ can be identified with the
Debye screening mass and the parameter $m_f$ can be identified as the
induced finite-temperature electron mass.
HTLpt is defined by treating
$\delta$ as a formal expansion parameter and expanding order by order
in $\delta$ around $\delta=1$.  This generates loops with fully dressed propagators and
vertices and also automatically generates
the counterterms necessary to remove the dressing as one proceeds to 
higher loop orders.

If the expansion in $\delta$ could be calculated to all orders,
the final result would not depend on $m_D$ or $m_f$.
However, any truncation of the expansion in $\delta$ produces results
that depend on $m_D$ and $m_f$.
Some prescription is required to determine $m_D$ and $m_f$
as a function of $T$ and $e$.  For example,
one can choose to treat both as variational parameters that should be
determined by minimizing the free energy or one can fix $m_D$ and
$m_f$ using a perturbative prescription.  We will compare both methods.
We will obtain the thermodynamic potential 
$\Omega(T,e,m_D,m_f,\mu,\delta=1)$ which is a function of the
mass parameters $m_D$ and $m_f$.  The free energy 
${\cal F}$
is obtained by evaluating the thermodynamic potential at the appropriate
values of the thermal masses.  Other thermodynamic 
functions can then be
obtained by taking appropriate derivatives of ${\cal F}$ with respect to $T$.

\section{Results}

In this section we present the final renormalized 
thermodynamic potential explicitly through order
$\delta^2$, aka NNLO, as obtained by us in Ref~\cite{Andersen:2009tw}.  
The final NNLO expression
is completely analytic; however, there are some numerically
determined constants which remain in the final expressions at
NLO.

\subsection{Next-to-leading order}

The renormalized NLO thermodynamic potential is
\begin{eqnarray}
\Omega_{\rm NLO}&=&
- {\pi^2 T^4\over45} \Bigg\{ 
	1 + {7\over4}N_f - 15 \hat m_D^3 
	- {45\over4}\left(\log\hat{\mu\over2}-{7\over2}+\gamma+{\pi^2\over3}
\right)\hat m_D^4
\nonumber \\ && %\hspace{-10mm}
	+ 60N_f\left(\pi^2-6\right)\hat m_f^4
	+ N_f{\alpha\over\pi} \Bigg[ -{25\over8}
	+ 15 \hat m_D
\nonumber \\ && %\hspace{-14mm}	
	+5\left(\log{\hat\mu \over 2}-2.33452\right)\hat m_D^2
	-45\left(\log{\hat\mu \over 2}+2.19581\right)\hat m_f^2
\nonumber \\ && %\hspace{-14mm}
	- 30\left(\log{\hat\mu \over 2}-{1\over2}
+\gamma+2\log2\right)\!\!\hat m_D^3
	+ 180\hat m_D \hat m_f^2 \Bigg]
\Bigg\} \;,
\label{Omega-NLO}
\end{eqnarray}
where we have introduced the dimensionless parameters
$\hat m_D = m_D /( 2 \pi T)$, $\hat m_f = m_f /( 2 \pi T)$ ,
and $\hat\mu = \mu /( 2 \pi T)$.

\subsection{Next-to-next-to-leading order}

The resulting NNLO thermodynamic potential is
\begin{eqnarray}\hspace{-40mm}
\Omega_{\rm NNLO}&=&
- {\pi^2 T^4\over45} \Bigg\{ 
	1 + {7\over4}N_f - {15\over4} \hat m_D^3 
	+ N_f {\alpha\over\pi} \Bigg[ -{25\over8}
	+ {15\over2} \hat m_D
\nonumber \\ &&\hspace{15mm}	
	+15 \left(\log{\hat\mu \over 2}
-{1\over2}+\gamma+2\log2\right)\!\hat m_D^3
	- 90\hat m_D \hat m_f^2 \Bigg]
\nonumber \\ &&\hspace{15mm}
+ N_f \left({\alpha\over\pi}\right)^2 
\Bigg[{15\over64}(35-32\log2)-{45\over2} \hat m_D\Bigg] 
\nonumber \\ &&\hspace{15mm}
+ N_f^2 \left({\alpha\over\pi}\right)^2 \Bigg[{25\over12}
\left(\log{\hat\mu \over 2}+{1\over20}+{3\over5}\gamma
\right. \nonumber \\ &&\hspace{20mm}	\left.
-{66\over25}\log2
+{4\over5}{\zeta^{\prime}(-1)\over\zeta(-1)}
-{2\over5}{\zeta^{\prime}(-3)\over\zeta(-3)}
\right)
\nonumber \\ &&\hspace{20mm}
+{5\over4}{1\over\hat m_D} - 15\left(\log{\hat\mu \over 2}-{1\over2}
+\gamma+2\log2\right)\!\hat m_D
\nonumber \\ &&\hspace{20mm}
+{30}{\hat{m}_f^2\over\hat{m}_D}
\Bigg]
\Bigg\} \;.
\label{Omega-NNLO}
\end{eqnarray}

\subsection{Free Energy}

\begin{figure}[t]
\begin{center}
\vspace{8mm}
\includegraphics[width=6cm]{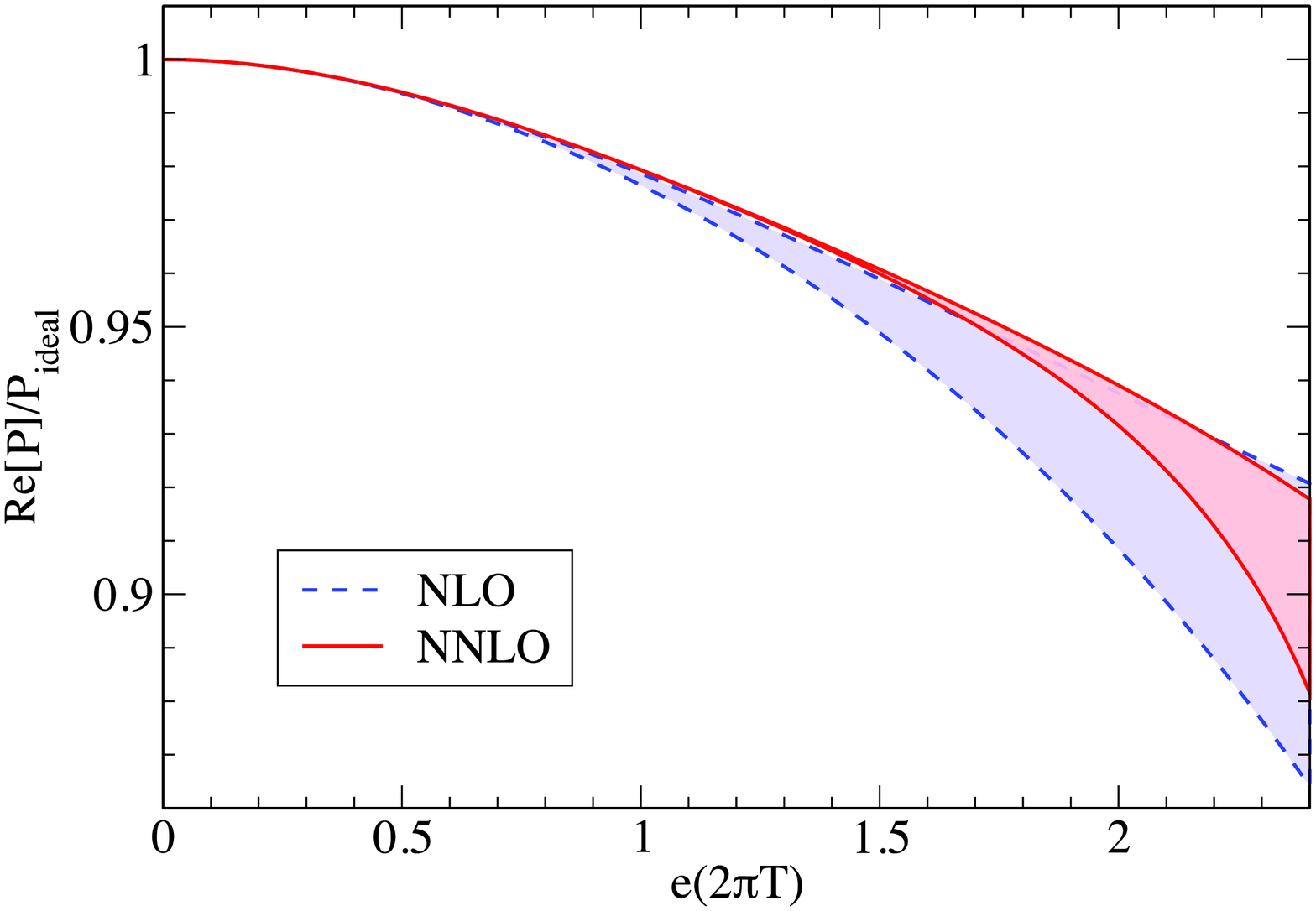}$\;\;\;\;$\includegraphics[width=6cm]{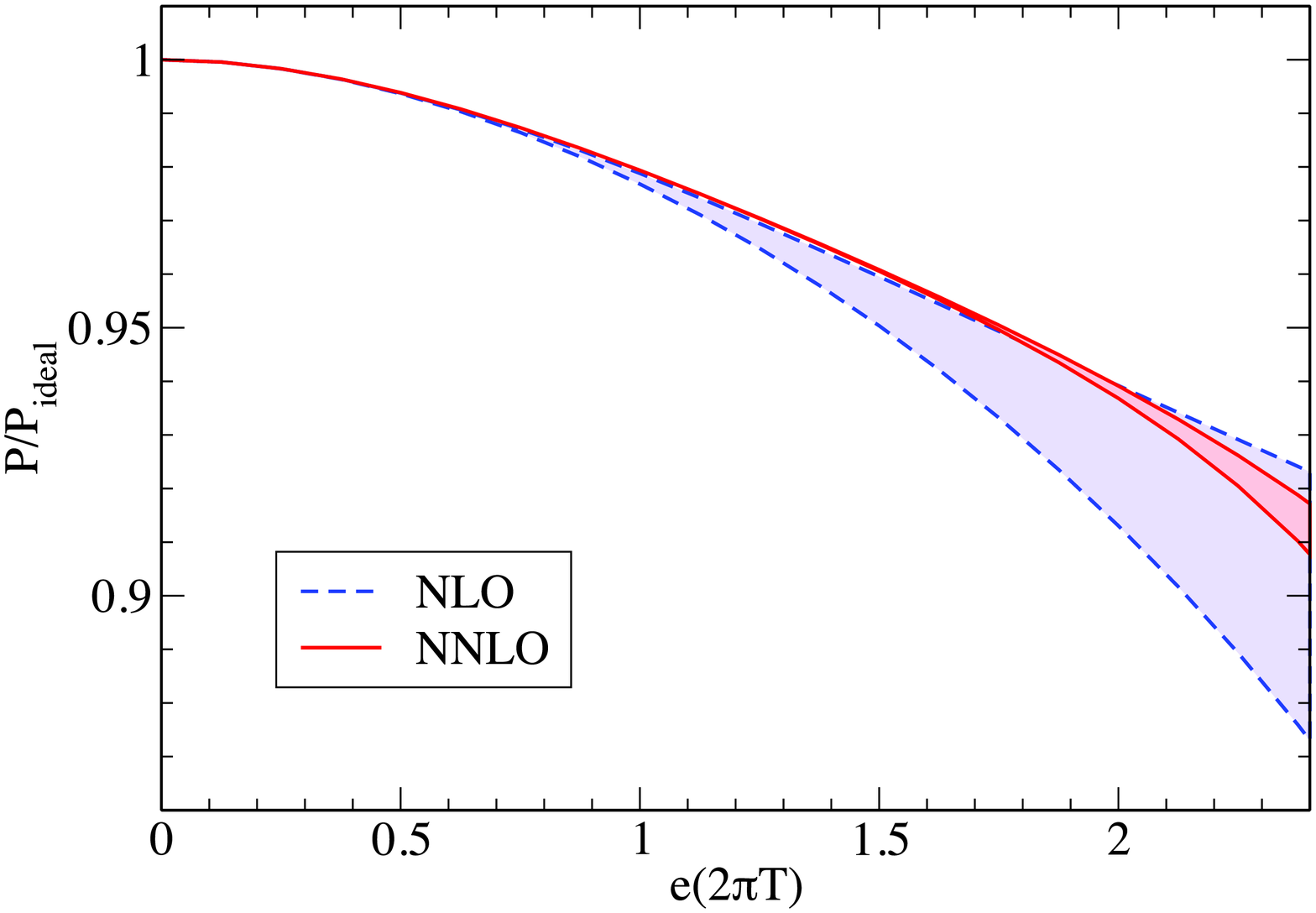}
\end{center}
%\vspace{-6mm}
\caption{A comparison of the renormalization scale variations between NLO and 
NNLO HTLpt predictions for the free energy of QED with $N_f=1$ and the 
variational Debye mass (left) and using the 
perturbative thermal masses (right). The bands correspond to varying the renormalization 
scale $\mu$ by a factor of 2 around $\mu=2\pi T$.}
\label{fig:NLONNLO}
\end{figure}

\begin{figure}[t]
\begin{center}
\vspace{6mm}
\includegraphics[width=9.5cm]{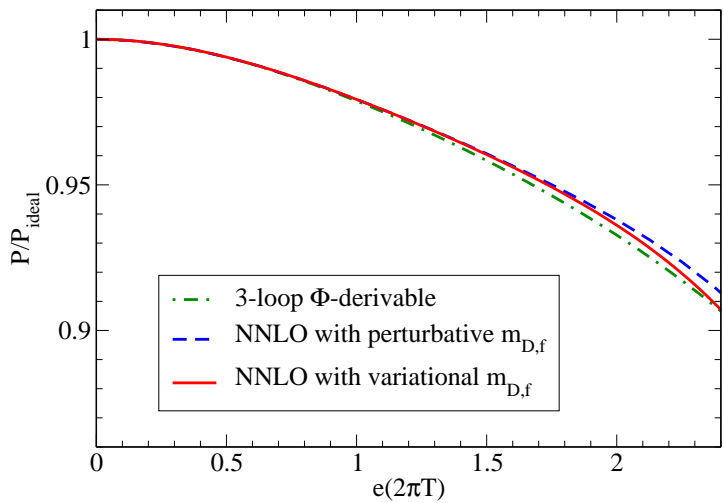}
\end{center}
\caption{A comparison of the predictions for the free energy of QED with 
$N_f=1$ between three-loop $\Phi$-derivable approximation \cite{phijm} and NNLO 
HTLpt at $\mu=2\pi T$.}
\label{fig:PhivsNNLO}
\end{figure}

The mass parameters $m_D$ and $m_f$ in hard-thermal-loop 
perturbation theory are in principle completely arbitrary. To complete a 
calculation, it is necessary to specify $m_D$ and $m_f$ as 
functions of $e$ and $T$.  In Ref.~\cite{Andersen:2009tw} we
considered two possible mass prescriptions in order to see how much
the results vary given the two different assumptions.  
First we considered the variational solutions for the thermal masses
and second we considered using the
$e^5$ perturbative expansion of the Debye mass 
\cite{Blaizot:1995kg,Andersen} and the $e^3$
perturbative expansion of the fermion mass~\cite{carrington}.
The resulting predictions for the free energy are shown in Fig.~\ref{fig:NLONNLO}.
As can be seen from these figures both the variational and perturbative mass
prescriptions seem to be consistent when going from NLO to NNLO.  As
a further check of our results in Fig. \ref{fig:PhivsNNLO} we show a comparison of our NNLO HTLpt 
results 
with a three-loop calculation obtained previously using a truncated three-loop 
$\Phi$-derivable approximation
\cite{phijm}.  As can be seen from this 
figure, there is
very good agreement between the NNLO $\Phi$-derivable and HTLpt approaches 
out to large
coupling.

%%%%%%%%%%%%%%%%%%%%%%%%%%%%%%%%%%%%%%%%%%%%%%%%%%%%%%%%%%%%%%%%
%%%%%%%%%%%%%%%%%%%%%%%%%%%%%%%%%%%%%%%%%%%%%%%%%%%%%%%%%%%%%%%%

%%%%%%%%%%%%%%%%%%%%%%%%%%%%%%%%%%%%%%%%%%%%%%%%%%%%%%%%%%%%%%%%
%%%%%%%%%%%%%%%%%%%%%%%%%%%%%%%%%%%%%%%%%%%%%%%%%%%%%%%%%%%%%%%%

\subsection{Conclusions and Outlook}

In this paper we discussed reorganizing finite temperature perturbation theory using HTLpt 
in order to improve the convergence of successive perturbative approximations to the free energy of
QED.  We presented results of a recent three-loop HTLpt calculation of the pressure in QED
\cite{Andersen:2009tw} and showed that the HTLpt reorganization improves the convergence of the 
successive approximations to the QED free energy at intermediate
coupling, $e \sim 2$.  We studied two different mass prescriptions and showed that the results 
for the free energy using both prescriptions were the same
to an accuracy of 0.6\% at $e=2.4$.   We also
compared the HTLpt three-loop result with a three-loop 
$\Phi$-derivable approach \cite{phijm} and found agreement at the 
subpercentage level.  In closing, we mention that 
the HTLpt reorganization is gauge invariant by construction we were able to 
obtain a completely analytic result for the resummed QED thermodynamic 
potential at three loops.  This gives
us confidence to apply the method also to full QCD.

\section*{Acknowledgments}

N. S. was supported by the Frankfurt International Graduate School for Science. 
M. S. was supported in part 
by the Helmholtz International Center for FAIR Landesoffensive zur Entwicklung 
Wissenschaftlich-\"Okonomischer Exzellenz program.

\end{document}